# Spin-orbit torque-driven skyrmion dynamics revealed by time-resolved X-ray microscopy

**Authors:** Seonghoon Woo,[1†*] Kyung Mee Song,[1,2†] Hee-Sung Han,[3] Min-Seung Jung,[4] Mi-Young Im,[4,5] Ki-Suk Lee,[3] Kun Soo Song,[1] Peter Fischer,[6,7] Jung-Il Hong,[4*] Jun Woo Choi,[1,8] Byoung-Chul Min,[1,8] Hyun Cheol Koo,[1,9] and Joonyeon Chang[1,8]

**Affiliations:**

[1]Center for Spintronics, Korea Institute of Science and Technology, Seoul 02792, Korea

[2]Department of Physics, Sookmyung Women's University, Seoul 04130, Korea

[3]School of Materials Science and Engineering, Ulsan National Institute of Science and Technology, Ulsan 44919, Korea

[4]Department of Emerging Materials, Research Center for Emerging Materials, DGIST, Daegu 42988, Korea

[5]Center for X-ray Optics, Lawrence Berkeley National Laboratory, Berkeley, California, 94720, USA

[6]Materials Sciences Division, Lawrence Berkeley National Laboratory, Berkeley, California, 94720, USA

[7]Department of Physics, University of California, Santa Cruz, California 94056, USA

[8]Department of Nanomaterials Science and Engineering, Korea University of Science and Technology, Daejeon 34113, Korea

[9]KU-KIST Graduate School of Converging Science and Technology, Korea University, Seoul 02792, Korea

[†] These authors equally contributed to this work

*Correspondence to: shwoo_@kist.re.kr, jihong@dgist.ac.kr

**Magnetic skyrmions are topologically-protected spin textures with attractive properties suitable for high-density and low-power spintronic device applications. Much effort has been dedicated to understanding the dynamical behaviours of the magnetic skyrmions. However, experimental observation of the ultrafast dynamics of this chiral magnetic texture in real space, which is the hallmark of its quasiparticle nature, has so far remained elusive. Here, we report nanosecond-dynamics of a 100 nm-size magnetic skyrmion during a current pulse application, using a time-resolved pump-probe soft X-ray imaging technique. We demonstrate that distinct dynamic excitation states of magnetic skyrmions, triggered by current-induced spin-orbit torques, can be reliably tuned by changing the magnitude of spin-orbit torques. Our findings show that the dynamics of magnetic skyrmions can be controlled by the spin-orbit torque on the nanosecond time scale, which points to exciting opportunities for ultrafast and novel skyrmionic applications in the future.**

**Introduction**

Most magnetic materials show collinear magnetic ordering due to the large exchange interaction between neighbouring magnetic moments. However, provided the Dzyaloshinskii-Moriya interaction (DMI)[1,2] is strong enough to overcome the exchange interaction, magnetic spins tend to align in non-collinear fashion with fixed homochirality. Most notably, in structures where the broken inversion symmetry leads to sufficiently large DMI, non-trivial small cylindrical swirling spin structures, called magnetic skyrmions can be energetically stable[3–16]. Unlike magnetic bubble domains[17,18], magnetic skyrmions[3–16] exhibit fascinating behaviors such as lattice formation[4,6,7,14], emergent electrodynamics[19] and current-driven displacement at low current densities[9–11]. Recently, micromagnetic simulations have predicted[11] and experiments confirmed[13–16] that magnetic skyrmions can be created at room temperature in robust metallic thin film heterostructures such as Ta/CoFeB/TaO$_x$[13], Pt/Co/Ta, Pt/CoFeB/MgO[14], Pt/Co/Ir[15] and Pt/Co/MgO[16] in the presence of a strong interfacial DMI. In such structures, current-pulse induced skyrmion displacement was achieved by harnessing spin-orbit torques (SOTs)[20–22] from non-magnetic heavy metals such as Ta and Pt. The experimental realization of the current-induced skyrmion motion is particularly important for the development of skyrmion-based memory devices[11,23]. However, earlier studies were limited to the static position imaging of magnetic skyrmions before and after the current pulses. Direct observation of nanosecond dynamics of sub-100 nm sized skyrmions in real space during the current pulse applications is required, not only for a better physical understanding of current driven skyrmion motion, but also for the exploration of new dynamical behaviours[12] of skyrmions. This has remained elusive so far, due to the high frequency and high spatial resolution measurement regimes required for such observations that are not easily accessible in experiments.

Here we report direct observation of magnetic skyrmion dynamics induced by SOTs in a transition metal ferromagnet multilayer at room temperature. By utilizing a stroboscopic pump-probe technique in a full-field magnetic transmission soft X-ray microscope (MTXM), we image the development of time-dependent transformation of magnetic skyrmions while a nanosecond current pulse is being applied. First, we demonstrate the generation of skyrmions at zero magnetic field by applying bipolar electric current pulses. We then observe that the skyrmions show two distinct dynamic behaviours, the breathing and the translational excitation behaviours, depending on the magnitude of SOTs, which is directly proportional to the current pulse amplitude.

**Results**

**X-ray microscopy observation of domain textures**

The multilayer structure in our study consists of [Pt (3 nm) / $Co_4Fe_4B_2$ (0.8 nm) / MgO (1.5 nm)]$_{20}$ (hereafter Pt/CoFeB/MgO) with perpendicular magnetic anisotropy (PMA). Figure 1a shows the out-of-plane hysteresis loop measured by vibrating sample magnetometry (VSM) of the Pt/CoFeB/MgO multilayer, while the inset shows that of the Pt/CoFeB/MgO unit layer. Vanishing remnant magnetization in the multilayer structure is due to the formation of multi-domain states in the absence of magnetic field, which results from the strong de-magnetizing field and the small domain wall surface energy of large DMI structures (see Supplementary Fig. 1 and Supplementary Note 1 for details). This is confirmed by the MTXM magnetic domain images in Fig. 1b, where dark and bright contrast corresponds to down and up domains, respectively. Micromagnetic simulations show that the magnetic texture in each layer is coupled magneto-statically across the entire stack (see Supplementary Fig. 2 and Supplementary Note 2 for details), thus the magnetic contrast is significantly enhanced in the transmission X-ray measurement. Since the pre-requisite for the formation of the chiral Néel DWs is a large DMI, we first determine the DMI constant analytically by measuring the field-dependent domain size variation for both up- and down-oriented domains from the sequence of MTXM images in Fig. 1b (see Supplementary Fig. 1 and Supplementary Note 1 for details). Alternatively the DMI constant in the Pt/CoFeB/MgO structure was quantified by field-driven domain expansion experiments and spin-Hall effect efficiency measurements (see Supplementary Fig. 3, and Supplementary Fig. 4, Supplementary Note 3 and Supplementary Note 4 for details). Using these measurements, we found that the magnitude of DMI is approximately $|D|$=1.68 mJ m$^{-2}$, which is in good agreement with reported values for Pt/CoFeB interface[14,24] and significantly larger than the critical value required to stabilize Néel DWs[25], $|D_c| \approx (2\ln 2)\mu_0 M_s^2 t / \pi^2 \approx 0.11$ mJ m$^{-2}$. Moreover, by analyzing the asymmetry of the field driven DW expansion (see Supplementary Fig. 3 and Supplementary Note 3 for details), we find that the Pt/CoFeB/MgO structures have homochiral left-handed Néel walls stabilized by DMI. This is shown schematically in Fig. 1c.

**Zero-field electrical generation of magnetic skyrmions**

To study electric current-driven magnetic domain dynamics in these systems, the magnetic multilayer films were patterned into 2 μm-wide magnetic strips, with 5 μm-wide and 100 nm-thick Au electrodes, used to inject electric current pulses, deposited on the top of the strips. Figure 2a shows the schematics of the electrical set-up used for the current-induced measurements. We first demonstrate the electrical generation of magnetic skyrmions in the absence of magnetic field using static X-ray measurements. Scanning electron microscope (SEM) image of our magnetic strip is shown in Fig. 2b, and two representative areas are separately indicated within the image. MTXM magnetic domain images acquired at zero field are shown in Fig. 2c and 2d. Before the pulse injection, two representative areas (left-panel images in Fig. 2c and 2d) of the magnetic strip showed the characteristic labyrinth multi-domain state. 20 ns-long bipolar voltage pulses with an amplitude of $V_a$=2.5 V (where peak-to-peak voltage $V_{pp}$=5 V), which corresponds to the current density of $\sim |j_a| = 1.6 \times 10^{11}$ A m$^{-2}$, were then applied to the strip. The pulses were applied at a repetition rate of $f$=3.33 MHz for a time span of five seconds. The right-panel images in Fig. 2c and 2d show that the labyrinth domains with chiral domain walls are completely transformed into multiple circular domains after the pulse applications, which we later confirm to indeed be magnetic skyrmions from the observation of their current driven dynamics. In this system, chiral domain wall motion is induced by the SOT at the Pt/CoFeB interface which switches polarity with the current pulse direction[22,24]. Then bipolar pulses are expected to excite the system by driving the domain walls back and forth. At the same time, we confirm that the magnetic properties of our film are preserved even after the pulse-induced skyrmion generation as discussed in Supplementary Fig. 5 and Supplementary Note 5. Therefore, such excitations could provide enough activation energy for the system to find the ground state magnetic domains without changing its magnetic properties (e.g. the net magnetization). Therefore, we believe that a stable multiple skyrmionic state of the Pt/CoFeB/MgO multilayer system can be achieved with a bipolar pulse-induced excitation. Note that the DMI constant of our system, $|D|$=1.68 mJ m$^{-2}$, is less than the threshold DMI value for the spontaneous skyrmion generation, $|D_{th}| = 2.26$ mJ m$^{-2}$. While a DMI larger than this threshold would lead to a negative domain wall energy ($\sigma_{DW} < 0$) thus resulting in an as-grown skyrmion state, it is clear that the relatively low DMI value of our material system cannot drive the system into multiple skyrmions or skyrmion lattice in the absence of external excitation. However, as the DMI value is still significantly larger than the critical value required stabilizing Néel DWs, $|D_c| = 0.11$ mJ m$^{-2}$, once the system reaches to a state with multiple skyrmions, the multi-skyrmion state can remain stable, i.e. at least a local energy minimum. The random size distribution and the lack of lattice structure of the skyrmions observed in Fig. 2b can also be explained by the DMI value lower than the threshold. Nevertheless, the observed electric current pulse-induced skyrmion generation is supported by the micromagnetic simulation study in Refs. [[12,26]] that show a current injection can provide sufficient energy to induce strong topological fluctuations to switch the system into a more stable state. Specifically, Ref. [[26]] describes the evolution of a single skyrmion state (net topological number, Q=1) from ferromagnetic state (Q=0) in a confined magnetic disk structure by nanoseconds bipolar-pulse injections with micromagnetic simulations, and shows that current pulse polarity-changes in nanoseconds time scale can induce the abrupt change of topological number, resulting in the emergence of topological state, skyrmions, after the pulse. A similar experimental observation was also reported in Ref. [14] in which skyrmions were stabilized after applying bipolar magnetic field oscillations.

**Time-resolved X-ray microscopy measurement**

Finally, we demonstrate current pulse-driven dynamics of skyrmions utilizing the SOTs from the spin Hall effect (SHE). It has been predicted that skyrmions may be nucleated and stabilized dynamically by the combination of vertical spin current injection and the corresponding spin transfer torque, in material systems with relatively large DMI[12]. Experimentally, this spin torque can be achieved by SOTs via SHE[22] from an in-plane current flowing along a non-magnetic heavy metal layer such as Pt. The spin Hall angle in our Pt/CoFeB/MgO structure is measured to be $\theta_{SH}$=+0.08, which is determined by the current-induced hysteresis loop shift experiments[24] (see Supplementary Note 4 for details). The large SOT originating from this SHE could efficiently actuate skyrmion motion in our magnetic track[11,13,14]. To study the skyrmion dynamics, time-resolved pump-probe MTXM experiments using the 2-bunch mode at Advanced Light Source (ALS) were performed (See Methods for details). Figure 3a shows the schematics of the device configuration and measurement scheme with an X-ray image of the device enclosed. Fig. 3a also shows an MTXM image of a single isolated skyrmion in this device. The isolated skyrmion state, generated by applying an external magnetic field $B_z$=+12.5 mT to the multiple skyrmion state, was chosen for the dynamic measurements to avoid possible skyrmion-skyrmion interactions. The actual temporal evolution of the bipolar current pulse is plotted in Fig. 3b, showing a rise/fall time of ~2.5 ns and an effective pulse width of roughly ~5 ns for each up/down pulse. Small reflected pulses are also observed due to the impedance mismatch between $R_{sample}$~100 Ω and $R_{ideal}$~50 Ω. For the time-resolved measurements, the amplitude of the bipolar current pulse was varied between $V_a$=1.5 V and $V_a$=2.5 V, which correspond to the current density of $|j_a| = 9.8 \times 10^{10}$ A m$^{-2}$ and $|j_a| = 1.6 \times 10^{11}$ A m$^{-2}$, respectively. When we first applied a low voltage of $V_a$=1.5 V to the sample, there was no observable change in shape as well as position of the skyrmion (see Supplementary Fig. 6 and Supplementary Note 6 for details). Then the pulse amplitude was increased. The dynamic behaviours of skyrmions upon the applications of pulses with $V_a$=2 V and 2.5 V are shown in Fig. 3c and 3d, respectively. The coloured circles in this plot indicate the time delay, which is also shown in Fig. 3b. When the pulse amplitude $V_a$=2 V (Fig. 3c), a strong breathing-like excitation behaviour of the skyrmion is observed. The center position of the skyrmion remains stationary at all times, while the skyrmion diameter changed by roughly a factor of two: from 190 nm at $t$=6 ns (positive pulse) to 85 nm at $t$=16 ns (negative pulse). The size variation was clearly noticeable due to the high spatial-resolution of the X-ray microcopy with the limit of ~25 nm due to the optical instruments. Details about the measurement of the skyrmion size are explained in Supplementary Fig. 7 and Supplementary Note 7. This observation of magnetic skyrmion breathing-like behaviour offers the first experimental demonstration of SOT-induced nanosecond dynamics for a magnetic skyrmion. Fig. 3e simultaneously shows time-dependent progression of skyrmion diameter and current pulse application, revealing the time-dependent skyrmion response to the voltage pulse. It is obvious that the skyrmion expands by ~260% of its initial area under a positive current pulse while it only shrinks to ~57 % of its initial area under a negative pulse. This asymmetric breathing-like excitation behaviour can be understood using the SOT+Oersted field model, which is described later in detail in Figure 4. Moreover, it should be noted that negligible time delay exists (less than a nanosecond measured from graph); the skyrmion size is largest when the pulse position is at its highest amplitude. This may indicate that the magnetic skyrmion in our structure has small inertia, whereas the inertia of non-chiral bubbles reported in Ref. [27] are rather large. As we further increase the pulse amplitude (Fig. 3e), a local translation of magnetic skyrmion is observed with the maximum displacement of ~84 nm, in addition to the skyrmion size variation. In Fig. 3f, we plot the time-dependent skyrmion displacement and pulse progression simultaneously, showing that a skyrmion moves upward with the velocity of ~10 m s$^{-1}$, which has only been shown statically in earlier studies[13,14]. It moves back to its original position during the negative cycle of the bipolar pulse. Therefore, temporal images in Fig. 3d effectively show the time-dependent displacement of a skyrmion during its current-induced translation. Fig. 3f can also be used to analyze inertial motion of a magnetic skyrmion. More thorough discussion on the inertial effect of skyrmions is discussed in the last part of this article. We also find that the skyrmion moves against the electron flow direction in our structure (schematically shown in Fig. 3d), as expected from the SOT-driven transport characteristics of left-handed Néel skyrmions[14,28]. The velocity and threshold for translational mode can be compared with a previous experiment on the static displacement of skyrmions in a similar Pt/CoFeB/MgO multilayer structure in Ref. [14], where the rough threshold for skyrmion motion $|j_c|=1.5\times10^{11}$ A m$^{-2}$ falls between the current densities applied in our study for Fig. 3c, $|j_a|=1.3\times10^{11}$ A m$^{-2}$, and Fig. 3d, $|j_a|=1.6\times10^{11}$ A m$^{-2}$; the velocity at similar current density also resembles such. From these experimental observations, we first speculate that SOTs could induce two different dynamical behaviours: the breathing-like and the translational excitation behaviours. Simply changing the external current pulse amplitude can tune between the modes. The full movie of the skyrmion motion can be found in Supplementary Movies 1-3.

**Micromagnetic simulation on skyrmion dynamics**

To support our experimental observation, we performed micromagnetic simulations shown in Figure 4 to further investigate the observed skyrmion dynamics (see Methods for simulation details). A larger magnetic field of $B_z$=+48 mT was applied for the simulation, compared to $B_z$=+12.5 mT used in experiments, to achieve a skyrmion size comparable with experiments and also avoid possible inter-skyrmion interaction in a given mesh-dimension with a periodic boundary condition. Fig. 4a first shows a skyrmion stabilized at its equilibrium state in the presence of external magnetic field, and Fig. 4b indicates the pulse profile used for the simulation, exhibiting a slightly different pulse shape from the one used in experiment (shown in Fig. 3b). In the computational approach, there is no reflected pulse because perfect impedance matching is assumed. First, to understand breathing-like excitation behaviour shown in Fig. 3c, we analyze time-dependent skyrmion diameter variation during the pulse application with three different models. The bipolar pulse amplitudes of $V_{\text{Low}}$ (=2 V) and $V_{\text{High}}$ (=6 V) are used. The simulation result presented in Fig. 4c first reveals that the application of SOTs, which induces strong topological fluctuation, only increases the skyrmion diameter regardless of pulse direction.

Thus, if our experimental observation is only driven by SOTs, we should have observed skyrmion breathing-like behaviour with a frequency, ~0.2 GHz, which is twice the pulse frequency. However, as observed in Fig. 3c and 3e, the breathing-like behaviour has the same frequency with the bipolar pulse, ~0.1 GHz, and the breathing amplitude is significantly larger with a noticeable asymmetry between expansion and contraction. This discrepancy can be compensated when we consider the Oersted field effect, which varied in linear proportion with current pulses. Note that our skyrmion is located near an Au electrode, as shown in the inset micrograph of Fig. 3a, thus, a finite Oersted field effect can be reasonably expected. Fig. 4c then shows the skyrmion size variation when we only consider the time-varying Oersted field generated by current pulses. A symmetric skyrmion breathing-like behaviour, driven by a bipolar Oersted field of 2 Oe per 1 V pulse amplitude, is observed. This model still fails to match the experimentally observed asymmetric expansion shown in Fig. 3c and 3e. However, as we combine these two effects simultaneously as shown in Fig. 4c, strong asymmetric breathing-like behaviour is achieved, and surprisingly, our experimental observation presented in Fig. 3c is reproduced qualitatively, as shown in Fig. 4d. Therefore, the suggested SOT+Oersted field model can successfully explain the experimentally observed skyrmion breathing-like behaviour. It should be noted that the observed breathing-like behaviour is not the internal skyrmion breathing mode, which happens when an oscillating magnetic field $h_z(t)$ matches the specific resonant criteria of surrounding Néel domain walls in the frequency regime of at least a few GHz as studied in Ref. [29]. Nevertheless, through the analysis shown in Fig. 4, we clearly show that the application of nanosecond current pulses induces a breathing-like skyrmion behaviour. In particular the simulation results in Fig. 4c suggest that SOT-only can still induce the size variation via strong topological fluctuation. In this case, an applied pulse would enlarge the skyrmion regardless of the pulse direction, and as soon as the pulse turns off, the skyrmion returns back to its equilibrium size. The breathing frequency can be easily tuned by the pulse-length. The SOT-driven breathing behaviour can be further maximized by using heavy metal layers with large SHE such as W ($|\theta_{SH}|$=0.3)[30] or facet-engineered IrMn$_3$ ($|\theta_{SH}|$=0.35)[31], since the SOT-driven size variation is linearly proportional to the SHE ($|\theta_{SH}|$=0.08 in our system). Therefore, we believe our result provides very important implication that, in an engineered material with strong spin orbit effect, we can realize a strong SOT-driven skyrmion breathing-like dynamics by simply turning on and off the external current pulses in the nanosecond time scale.

We then applied the same models for analysing skyrmion travel distance as a function of time, as shown in Fig. 4e and 4f. Fig. 4e clearly show that the SOT can induce translational motion of the skyrmions, while the Oersted field cannot. When the experimental result (Fig. 3f) and computed model (Fig. 4f) are compared, the time-dependent travel distances agree qualitatively. However, unlike the experiments in which the translational skyrmion mode was only observed for current pulses larger than the threshold, in simulations, the pulse-injection always triggers a simultaneous skyrmion displacement as shown in Fig. 4e. This difference could be explained by material deficiencies, such as non-uniform DMI, which were not considered in simulations. In non-epitaxial sputtered films, the DMI constant can be non-uniform over a single film, resulting in a finite pinning threshold that traps skyrmion and requires large current to actuate its motion[14]. Therefore, it is likely that there exists sizable DMI variation over the actual sample, which effectively provides pinning threshold, so that only radial size variation was observed in Fig. 3c without translational motion. Once the excitation energy overcomes the threshold with larger current pulses, the translational motion is actuated. Considering the ever-present imperfections in real magnetic films, we believe our experimental observations provide a practical way to induce and tune between distinct skyrmion dynamic states simply by changing the SOTs. However, we emphasize that our observed skyrmion was not located at a strong local pinning site as discussed in Supplementary Fig. 8 and Supplementary Note 8. Finally, it should be noted that, while the experimental and computational results of the skyrmion size variation and travel distance agree qualitatively, a quantitative difference persists, and additional investigations will be pursued in further studies.

**Inertial effect of magnetic skyrmions**

The time-dependent skyrmion displacement, shown in Fig. 4e, could be used for studying the inertial motion of magnetic skyrmions. Based on the successful agreement between experiment and simulation (Fig. 4f), the computational data in Fig. 4f is utilized for more in-depth analysis. We first plot the travel distance and voltage pulse as a function of time in Figure 5a, and also plot its derivative, velocity, and voltage pulse in Fig. 5b. In spin textures such as skyrmions and domain walls with massless-particle approximation, as can be derived from the Thiele's equation[32], the velocity is linearly proportional to external forces. Thus, if there exists an inertial effect, we should be able to observe a finite skyrmion displacement, implying finite velocity, even in the absence of external force, which is the current pulse in the present case. However, as we plot the velocity, $v = \frac{dD}{dt}$, as a function of time in Fig. 5b, it is evident that the velocity and force curves are perfectly matching, implying that there is no observable inertial effect on the skyrmion motion in our system, up to the spatial resolution limit ~25 nm of the MTXM. The agreement between the computational model and the time-dependent displacement of experimentally observed skyrmions further validates our conclusion on inertial motion even for the real case. According to the Landau-Lifshitiz-Gilbert (LLG) equation as a consequence of the first order derivative in time, each spin in a texture such as a domain wall or a skyrmion should simultaneously obtain their velocity from any external forces such as magnetic field and/or electrical current, and this is clearly shown in Ref. [33].

**Discussion**

However, it should be pointed out that our observation disagrees with the recent experimental observation of skyrmion bubble dynamics, showing the presence of strong inertia and the areal mass being as large as $2.0 \times 10^{-7}$ kg m$^{-2}$.[27] We believe the difference originates from the fact that the inertial motion of a skyrmion can be enhanced in a confined geometry. In a situation where spin dynamics can be fully reproduced using the LLG equation, inertial behaviour cannot appear as we stated above. Indeed, it has been shown that the dynamics of vortices in thin ferromagnetic films can be successfully reproduced without introducing the concept of inertia, where the vortex state has a finite topological invariant known as the skyrmion charge $q = (1/4\pi) \int m \cdot (\partial_x m \times \partial_y m) dx\,dy$; a vortex has $q = \pm 1/2$.[34,35] However, in situations where the LLG equation fails so that the collective coordinate should be implemented, the concept of inertia can be introduced to correctly reproduce the observed spin dynamics such as skyrmion bubble gyration in a confined geometry.[27,36] In our case where a skyrmion is located in a quasi-infinite thin film, its dynamic behaviour such as translation or breathing-like motion can be described without implementing the collective coordinate model. During the submission of this work, Litzius *et al.*[37] reported the observation of skyrmion Hall effect in Pt/CoFeB/MgO multilayers using time-resolved X-ray microscopy method, which works in very similar fashion with MTXM used in this work.

In summary, we investigated the ultrafast dynamics of a magnetic skyrmion in the nanosecond time scale using time-resolved X-ray microscopy in order to better understand the physical origin behind the current induced skyrmionic motion. We demonstrated the electrical generation of magnetic skyrmions in the absence of magnetic field by using bipolar pulses. By changing the magnitude of SHE-induced SOTs with the current amplitude, we have revealed the manipulation of skyrmion dynamics between the breathing-like and the translational excitation behaviors. Our observations offer an efficient method to manipulate magnetic skyrmions both statically and dynamically, which are an essential requirement for the development of broadband low-power and high-density skyrmionic devices. Therefore, our observation indeed opens the door to versatile and novel skyrmionic applications.

**Methods**

**Sample preparation and experimental method**

The [Pt(3 nm)/CoFeB(0.8 nm)/MgO(1.5 nm)]$_{20}$ films were grown on a 100nm-thick SiN substrate by dc magnetron sputter deposition at room temperature under 3 mTorr Ar for Pt, 1 mTorr Ar for CoFeB and 4 mTorr Ar for MgO, with a background pressure lower than ~3x10$^{-8}$ Torr. Ta seed, $t_{seed}$=3 nm, and capping, $t_{capping}$=2 nm, layers were deposited for better adhesion to SiN substrate and protection from atmospheric conditions, respectively. The thin film samples then patterned using electron beam lithography and ion-milling techniques. Nominally identical films were grown on SiO$_x$/Si substrates for vibrating sample magnetometry (VSM) measurements. The VSM measurements yielded an anisotropy field of $\mu_0 H_k$=0.7 T, and a saturation magnetization of $M_s$=9x10$^5$ A/m. The current contacts in Fig. 2 and Fig. 3 consisted of Ti(5 nm)/Au(100 nm), which was deposited using the same dc magnetron sputtering. The contacts were also patterned using electron beam lithography and lift-off. For the current-density calculations used in Fig. 3 and maintext, we only considered Pt and CoFeB layers while excluding Ta layers, which is expected to be high resistivity $\beta$-phase.

All images in the main text and supplementary materials were acquired using full-field MTXM performed at the XM-1 beamline 6.1.2 at the Advanced Light Source (ALS) in Berkeley, California. The device used for experiments was 2 μm-wide with electrical resistance of ~100 Ω between two Au contacts. Pulse current densities above ~2x10$^{11}$ A m$^{-2}$ led to the damage of the Au contact, which eventually limited the maximum current applied in Fig. 3. In time-resolved 2-bunch experiments, where X-rays are injected at 300 ns intervals (injection frequency of 3.33 MHz), the stroboscopic pump-probe technique of MTXM restricts the imaging to fully reproducible magnetic events by synchronizing the incoming X-ray photon flashes (probe) and injecting current pulses (pump). The time evolution of the dynamics was recorded by varying the delay times between the photon flashes and the excitation pulses. Due to the low intensity for a single X-ray pulse, about $10^8$-$10^9$ pump-probe events are required to obtain a single magnetic image shown in Fig. 3, which corresponds to an accumulation time per image of a few tens of seconds. To ensure that all acquired images in Fig. 3 indicate the exactly same area of our ferromagnetic wire, we have performed the image-displacement correction by aligning Au contact boundary near our skyrmion across all acquired images using computer software. This allows us to align all images to the same position within error of a pixel size, ~25 nm.

**Simulation method**

Micromagnetic simulations were performed using open-source MuMax$^3$ and it solved Landau-Lifshitz-Gilbert (LLG) equation: $\partial \mathbf{m}/\partial t = -\gamma_0 \mathbf{m} \times \mathbf{H}_{eff} + \alpha \mathbf{m} \times \partial \mathbf{m}/\partial t - \gamma_0 \mathbf{m} \times (\mathbf{m} \times \mathbf{H}_{SH}\hat{y})$[38,39] where $H_{SH} = \mu_B \theta_{SH} j_a / \gamma_0 e M_s t_z$ with normalized local magnetization vector $\mathbf{m}$, the gyromagnetic ratio $\gamma_0$, the effective field vector $\mathbf{H}_{eff}$, damping constant $\alpha$, Bohr magneton $\mu_B$, spin Hall angle $\theta_{SH}$, the current density $j_a$ which flows in the sample along $x$-direction, saturation magnetization $M_s$, charge of electron $e$, and thickness of ferromagnetic layer $t_z$. To model multilayer film [Pt(3 nm)/ CoFeB(0.8 nm)/MgO(1.5 nm)]$_{20}$, we used an effective medium model with the cell size 2×2×0.8 nm$^3$ for the 1000×1000 nm$^2$ mesh.[14] To ignore the effect of the lateral sample edges, we adopt a periodic boundary condition along both in-plane directions. Material parameters used are an exchange stiffness $A_{ex}=1.4\times10^{-11}$ J m$^{-1}$, saturation magnetization $M_s=9\times10^5$ A m$^{-1}$, Dzyaloshinskii Moriya constant $D=1.68$ mJ m$^{-2}$, uniaxial anisotropy constant $K_u=7.79\times10^5$ A m$^{-3}$, damping constant $\alpha=0.5$, and spin Hall angle $\theta_{SH}=+0.08$. The current density $j_a=5.48\times10^{10}$ A m$^{-2}$ corresponds to $V_a=1$ V in SOT-related simulations. The amplitude of the Oersted field derived from the flowing current in Au electrode is considered to be 2 Oe (0.2 mT) per 1 V pulse amplitude.

**Data availability**

Data supporting the findings of this study are available within the artic and its Supplementary Information files and from the corresponding author upon request.

**Acknowledgments:**

Work at KIST was primarily supported by the KIST Institutional Program and the National Research Council of Science & Technology (NST) grant (No. CAP-16-01-KIST) by the Korea government (MSIP), and the Pioneer Research Center (2011-0027905). This work was also supported by the National Research Foundation of Korea(NRF) funded by the Ministry of Science, ICT and Future Planning (2016K1A3A7A09005418, 2012K1A4A3053565, 2015M3D1A1070465, 2014R1A2A2A01003709 and 2015R1C1A1A02037742). Supply of a schematic drawing shown in Fig. 1c from Dr. Ki-Young Lee is gratefully acknowledged. S.W. acknowledges the support from the POSCO Science Fellowship of POSCO TJ Park Foundation. K.M.S acknowledges the support from the Sookmyung Women's University BK21 Plus Scholarship. P.F. acknowledges support from the Director, Office of Science, Office of Basic Energy Sciences, Materials Sciences and Engineering Division, of the U.S. Department of

**Author Contributions**

S.W. designed and initiated the study. S.W. and K.M.S. fabricated devices and performed the film characterization. K.S.S. provided technical input for nano-fabrication using e-beam lithography. S.W., K.M.S., M.-S.J., M.-Y.I. and J.W.C. performed X-ray imaging experiments using MTXM with supports from J.-I.H. at the Advanced Light Source in Berkeley, California. H.-S.H. and K.-S.L. performed micromagnetic simulations. All authors participated in the discussion and interpretation of the results. S.W. and J.W.C. drafted the manuscript and revised it with assistance from M.-Y.I., K.-S.L., P.F., J.-I.H., B.-C., H.C.K. and J.C. All authors commented on the manuscript.

**Competing financial interests**

The authors declare no competing financial interests.

**Author Information**

S.W. and K.M.S. contributed equally to this work. Correspondence and requests for materials should be addressed to S.W. (shwoo_@kist.re.kr) and J.-I.H. (jihong@dgist.ac.kr)

**Figure legends.**

**Figure 1. Structural characteristics and X-ray imaging of domain patterns.** (**a**) Out-of-plane hysteresis loop for a companion multilayer film grown on Si wafer. The magnetic moment is normalized to its saturation magnetization $M_S=9\times10^5$ A m$^{-1}$. The inset shows both in-plane and out-of-plane hysteresis loops for a Pt/CoFeB/MgO unit layer grown on Si wafer. (**b**) Series of magnetic transmission X-ray microscopy (MTXM) images acquired for increasing field $B_z > 0$. Dark and light contrast corresponds to down (-$z$) and up (+$z$)–oriented magnetization directions, respectively. Scale bar, 200 nm. (**c**) MTXM image acquired at $B_z=25$ mT and the schematic of the expected chiral magnetic textures.

**Figure 2. Electrical generation of magnetic skyrmions.** (**a**) Schematic of the electric connection for current pulse-injection experiments. The X-ray beam is injected at the frequency of 3.33 MHz (bunch spacing = 300 ns) and can be synchronized with current pulses to observe temporal evolution of magnetic textures. A source meter is connected through a bias-tee to simultaneously measure the device resistance. The scanning electron microscope (SEM) image of the actual device is also shown. Scale bar, 5 μm. (**b**) SEM image shows two distinct areas in a magnetic wire, (i) and (ii), where the magnetic textures were separately measured. Magnetic transmission X-ray microscopy (MTXM) images of the magnetic domain states in (**c**), region-(i), and (**d**), region-(ii), showing the transformation of domain phase from labyrinth state to multiple skyrmionic state after the application of bipolar pulses at $B_z=0$ mT. Enclosed Bipolar pulse trains are injected at $f=3.33$ MHz for 5 s. Images are taken before and after the current pulses.

**Figure 3. Dynamic behavior of a magnetic skyrmion induced by bipolar pulse injections.** (**a**) Schematic of the Pt/CoFeB/MgO magnetic wire on the Si$_3$N$_4$ membrane with electrode contacts. The skyrmions are stabilized in the magnetic wire at $B_z=12.5$ mT. The inset shows an initial skyrmion used for the dynamic measurement. (**b**) Pulse profile used for the dynamics measurement. There is a time delay between each bipolar pulse and the incident X-ray beam pulse, which are injected at a frequency of $f=3.33$ MHz. The coloured circles and time numbers, t1~t10, in this plot indicate the time delay, which is also shown in following images. Magnetic skyrmion configuration at different time delays for voltage amplitudes of (**c**) $V_a =2$ V and (**d**) $V_a=2.5$ V, respectively. Arrows are included in (**e**) to indicate the direction and magnitude of electron flow at each time delay, and a schematic drawing is also included on the right side to more clearly show the skyrmion motion direction relative to the electron flow direction. Horizontal lines are drawn in (**c**) and (**d**) to more effectively show the variation of the skyrmion. Scale bar, 200nm. (**e**) Measured skyrmion diameter and (**f**) total skyrmion travel distance as a function of pulse delay time, respectively. The actual pulse profile, shown in (**b**), is also included in both plots to effectively show the response of skyrmion to time-dependent pulses. Error bars in (**e**) and (**f**) indicate the spatial resolution of beamline, ~25 nm.

**Figure 4. Micromagnetic simulations on the skyrmion dynamics triggered by spin orbit torques.** (**a**) Top-view of a simulated skyrmion structure at its equilibrium state when $B_z=48$ mT. Scale bar, 200 nm. (**b**) Pulse profile used for the simulation, which is slightly different from the actual profile used in experiments due to impedance mismatching in experiments. (**c**) Time-dependent skyrmion size variation when we consider spin orbit torque (SOT)-only, Oersted field-only and SOT and Oersted field simultaneously. (**d**) Experimentally measured (left-axis)

and computed (right-axis) skyrmion diameter as a function of pulse delay time. (**e**) Time-dependent skyrmion travel distance of SOT-only, Oersted field-only and SOT and Oersted field models. (**f**) Experimentally measured (left-axis) and computed (right-axis) skyrmion travel distance as a function of pulse delay time. Note that y-axis scales are different between the experiment and computational model in (**d**) and (**e**)**.** Error bars in (**d**) and (**f**) indicate the spatial resolution of beamline, ~25 nm.

**Figure 5. Micromagnetic simulations on the inertial motion of a skyrmion.** (**a**) Calculated time-dependent pulse profile and corresponding skyrmion travel distance. (**b**) First time-derivative of skyrmion travel distance, resulting in the velocity and pulse amplitude as a function of pulse delay time.

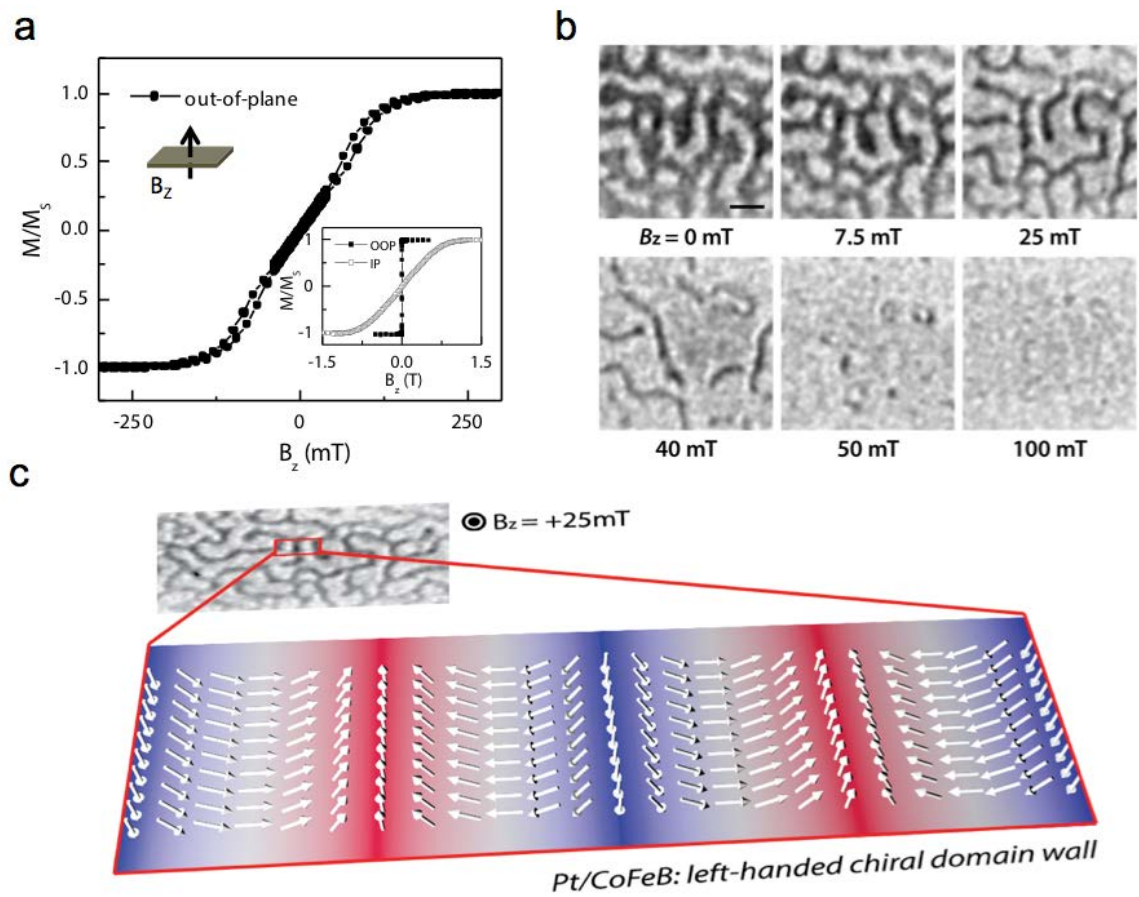

**Figure 1**

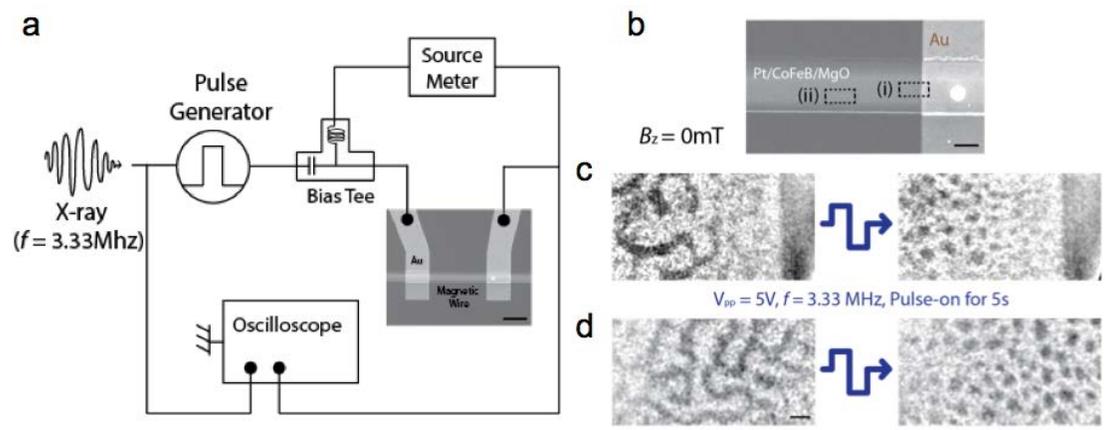

**Figure 2**

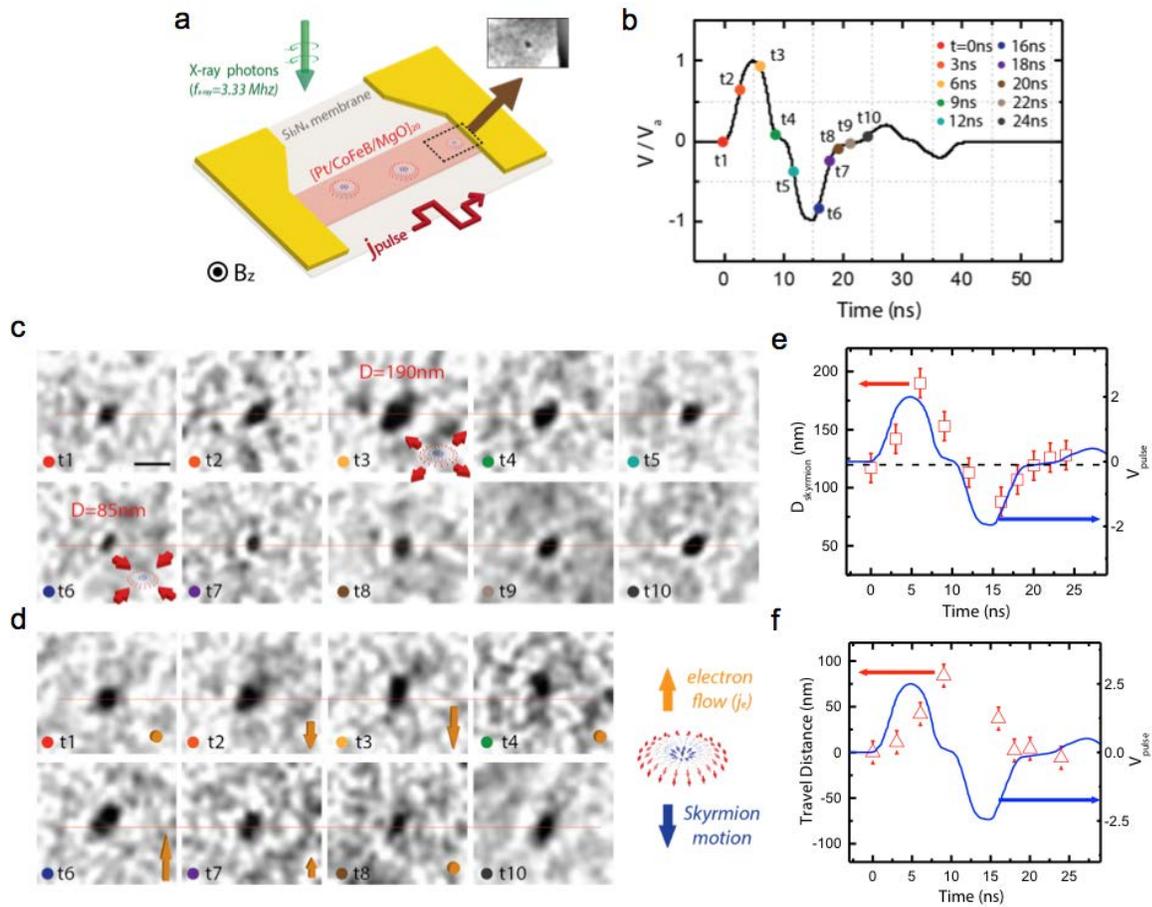

**Figure 3**

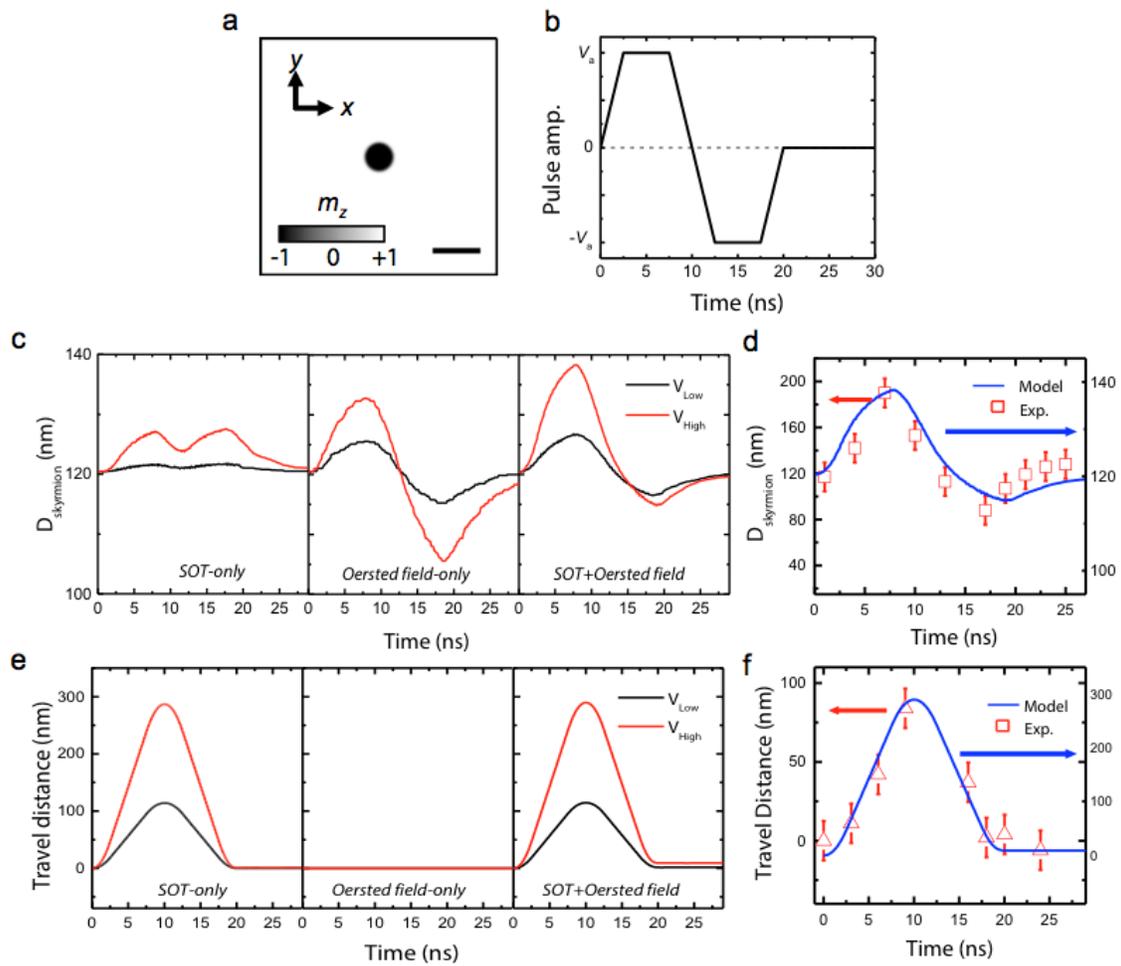

**Figure 4**

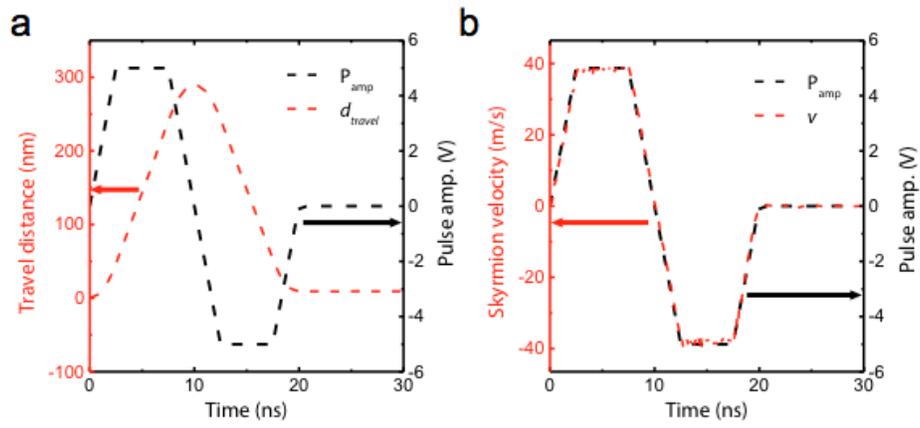

**Figure 5**